\documentclass[journal=ancac3,manuscript=article]{achemso}
\usepackage{lipsum}
\usepackage{graphicx}
\usepackage{amsmath}
\usepackage{amssymb}
\usepackage{subeqnarray}
\usepackage{color}
\usepackage{bbold}
\usepackage{dsfont}
\usepackage{tikz}
\usepackage{cancel}
\usetikzlibrary{positioning,calc}
\usepackage{float}
\usepackage[version=3]{mhchem}
\usepackage{color}
\usepackage[normalem]{ulem}

\newcommand{\eqn}[2]{\begin{equation}\label{#2}#1\end{equation}}
\newcommand{\mt}[1]{\mathrm{#1}}

\newcommand{\fig}[1]{Figure~\ref{#1}}

\newcommand{\E}[1]{\times 10^{#1}}

\newcommand{\angstrom}{\textup{\AA}}

\newcommand{\ie}{\textit{i.e.},~}
\newcommand{\via}{\textit{via~}}
\newcommand{\eg}{\textit{e.g.,~}}
\newcommand{\etal}{\textit{et al.~}}

\newcommand{\mum}{~$\mt{\mu m}$~}

\newcommand{\cm}{$\mt{cm^{-1}}$~}

\newcommand{\hfo}{\ce{HfO_x}~}
\newcommand{\tao}{\ce{TaO_x}~}

\newcommand{\vth}{\ensuremath{V_{TH}}~}
\newcommand{\ion}{\ensuremath{I_{ON}}~}
\graphicspath{{./Figures/}}

\author{Sahej Sharma}
\affiliation{School of Mechanical Engineering, Purdue University, West Lafayette, IN 47907, USA}
\alsoaffiliation{Birck Nanotechnology Center, Purdue University, West Lafayette, IN 47907, USA}
\author{Shao-Heng Yang}
\affiliation{School of Mechanical Engineering, Purdue University, West Lafayette, IN 47907, USA}
\alsoaffiliation{Birck Nanotechnology Center, Purdue University, West Lafayette, IN 47907, USA}
\author{Himani Jawa}
\affiliation{Elmore School of Electrical and Computer Engineering, Purdue University, West Lafayette, IN 47907, USA}
\alsoaffiliation{Birck Nanotechnology Center, Purdue University, West Lafayette, IN 47907, USA}
\author{Rana Yuvraj}
\affiliation{School of Mechanical Engineering, Purdue University, West Lafayette, IN 47907, USA}
\alsoaffiliation{Birck Nanotechnology Center, Purdue University, West Lafayette, IN 47907, USA}
\author{Bach Nguyen}
\affiliation{School of Mechanical Engineering, Purdue University, West Lafayette, IN 47907, USA}
\alsoaffiliation{Birck Nanotechnology Center, Purdue University, West Lafayette, IN 47907, USA}
\author{Chang Niu}
\author{Shiva Radhakrishnan}
\author{Shalini Tripathi}
\affiliation{imec USA, West Lafayette, Indiana 47901, USA}
\author{Dennis Lin}
\author{Cesar Javier Lockhart de la Rosa}
\author{Pierre Morin}
\affiliation{imec, Leuven 3001, Belgium}
\author{Dmitry Zemlyanov}
\affiliation{Birck Nanotechnology Center, Purdue University, West Lafayette, IN 47907, USA}
\author{Francesca Iacopi}
\affiliation{imec USA, West Lafayette, Indiana 47901, USA}
\author{Zhihong Chen}
\affiliation{School of Mechanical Engineering, Purdue University, West Lafayette, IN 47907, USA}
\alsoaffiliation{Birck Nanotechnology Center, Purdue University, West Lafayette, IN 47907, USA}
\author{Joerg Appenzeller}
\affiliation{Elmore School of Electrical and Computer Engineering, Purdue University, West Lafayette, IN 47907, USA}
\alsoaffiliation{Birck Nanotechnology Center, Purdue University, West Lafayette, IN 47907, USA}
\author{Thomas E. Beechem}
\email{tbeechem@purdue.edu}
\affiliation{School of Mechanical Engineering, Purdue University, West Lafayette, IN 47907, USA}
\alsoaffiliation{Birck Nanotechnology Center, Purdue University, West Lafayette, IN 47907, USA}
\date{\today}
\title{Seed Layer Engineering for Effective Charge Transfer Doping of \ce{MoS_2} Transistors}

\begin{document}
\begin{abstract}
Integrating two-dimensional semiconductors such as \ce{MoS_2} with dielectric materials remains a central challenge for their use in future logic technologies. While seed layers are typically introduced to promote dielectric nucleation and adhesion, we show that they also critically govern charge-transfer doping and, in turn, transistor performance. Back-gated monolayer \ce{MoS_2} transistors passivated on their top-surface with a Ta-seed/\ce{HfO_x} dielectric stack were fabricated and characterized electrically and physically using Raman, photoluminescence, and X-ray photoelectron spectroscopies. Threshold voltage and on-current varied strongly with Ta-seed thickness and deposition conditions, and these changes correlated with signatures observed across all spectroscopic probes. The results reveal that the seed layer both introduces disorder into the \ce{MoS_2} channel and modifies the interfacial charge environment controlling charge transfer between \ce{HfO_x} and \ce{MoS_2}. Optical spectroscopy shows that on-current tracks seed-induced disorder, whereas X-ray photoelectron spectroscopy indicates that threshold voltage correlates with shifts in the local electrostatic environment associated with interfacial charge transfer. Better performance was obtained with ultrathin 0.2 nm Ta seed layers deposited under oxygen-poor conditions, which limit deposition-induced damage while facilitating charge transfer. These findings identify seed-layer engineering as a key strategy for controlling disorder and interfacial doping in \ce{MoS_2} devices and establish multimodal spectroscopy as a practical during-fabrication approach for process development and monitoring. 
\end{abstract}

\begin{tocentry}
\centering
\includegraphics[width=3.25in,height=1.75in,keepaspectratio]{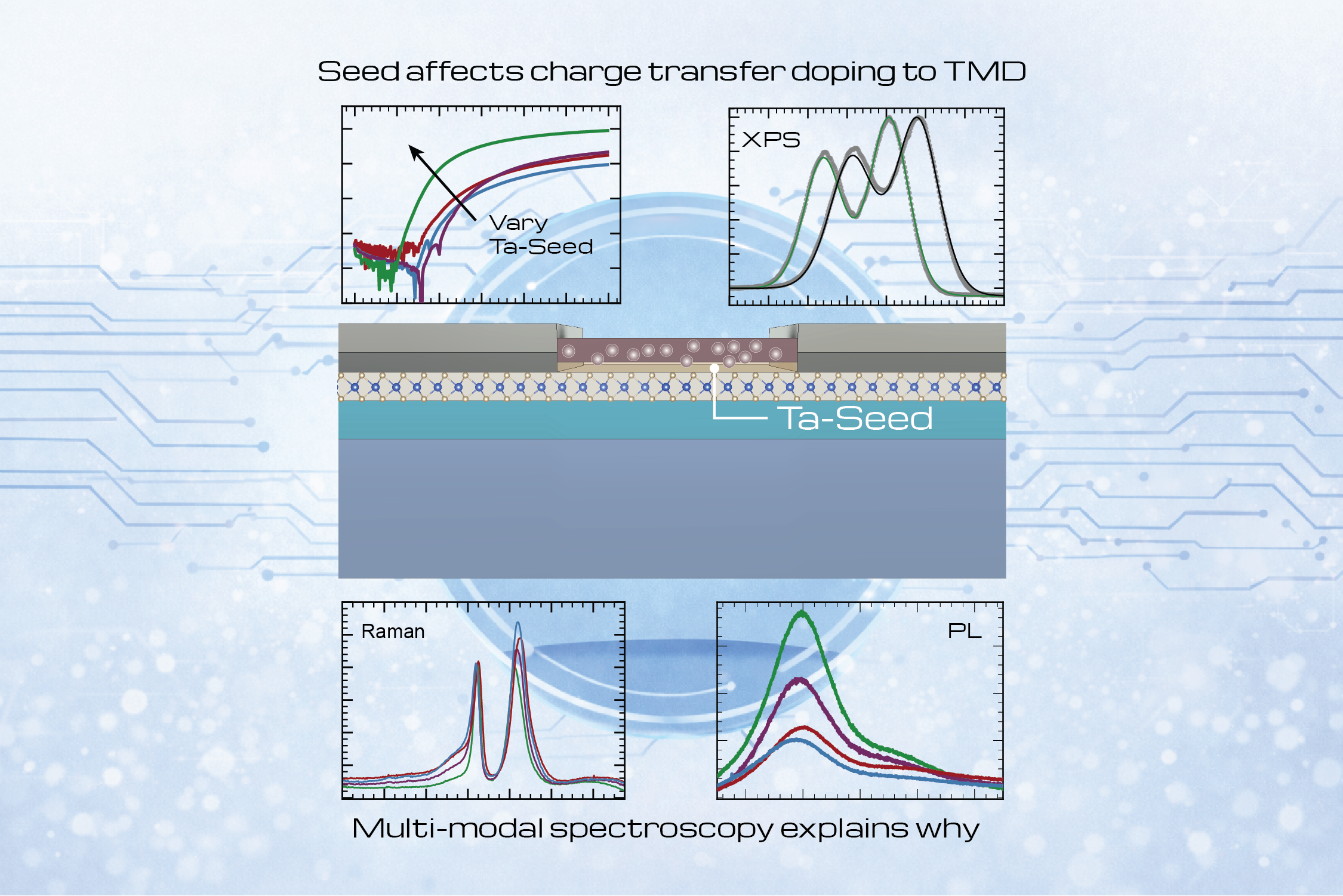}
\end{tocentry}

\maketitle

Continued ultrascaling of logic transistors now motivates changes to not only device architecture but even the channel material itself. \cite{kim_2025e} The transition to silicon nanosheet gate all-around (GAA) transistors is suggestive of a two-dimensional (2D) channel and thus materials that themselves are inherently 2D.  Along these lines, two-dimensional transition-metal dichalcogenides (TMD) like \ce{MoS_2} have attracted significant attention because of their atomically thin body thickness enabling improved gate control and scaling potential that comes inherently in ``losing" a dimension.\cite{huang_2022c, liu_2024c} These material advantages combined with experimental demonstrations of devices exhibiting high on-off ratios (\ce{I_{on}/I_{off}}) of $10^6$ with nearly ideal subthreshold swing (SS) of 65 mV/dec have motivated their pursuit for next-generation logic. \cite{liu_2024c, radisavljevic_2011} Despite the promise and investment, significant hurdles persist in reliably achieving the requisite device characteristics at scale.\cite{knobloch_2025} 

Among these hurdles, dielectric integration is one of the most significant.  The dielectric determines nearly every aspect of device performance.   Ideally, it enables deft gate control of charge density within the TMD, preserving the high carrier mobility in the channel, while exhibiting minimal leakage. Beyond electrostatic control, the dielectric is also particularly effective in doping the TMD channel \via so-called remote-, modulation-, or charge-transfer schemes.\cite{valsaraj_2015,illarionov_2020} Realized with several metal-oxides (\eg \ce{AlO_x},\cite{mcclellan_2021} \ce{TiO_x},\cite{rai_2015} \ce{TaO_x},\cite{lan_2023} \ce{HfO_x}\cite{ko_2025}), charge-transfer doping leverages strategic band alignment between the dielectric and the TMD to realize higher carrier concentrations apart from the necessity of physically placing a mobility reducing dopant in, or directly atop, the TMD. The remote nature of the doping, in turn, has led to some of the highest performing \ce{MoS_2} devices to date.\cite{mcclellan_2021,lan_2023}  

Realizing the gains of charge-transfer doping along with reasonable levels of subthreshold swing necessitates that the interface between the dielectric and the TMD be optimized as well. Creating ``good" interfaces with TMDs is not straightforward, however.  The difficulty is a consequence of their van Der Waals structure.  Unlike silicon, which forms a high-quality native semiconductor-oxide interface, the lack of surface sites on the TMD can not only limit adhesion but also result in dangling bonds in the dielectric film that create trap states causing increased subthreshold swing, hysteresis, and variability in \vth\!.\cite{yang_2023c, knobloch_2025}  These same interface traps can also ``steal" the charge that would otherwise dope the TMD channel \via charge-transfer. Furthermore, the deposition of the dielectric can itself damage the TMD lowering its mobility. 

A common approach to limit these non-ideal effects is the use of an ultrathin (\textit{c.a.} $\lesssim$1 nm) seed, or buffer, layer deposited prior to growth of the dielectric used for gating. Numerous materials have been investigated as possible seeding layers including oxides, organic buffers, and metals.\cite{wirtz_2015, shearer_2025}  Among these options, Ta-metal seeds oxidized prior to dielectric deposition have led to promising \ce{MoS_2} transistor characteristics.\cite{lan_2023} This is thought to be a consequence of the interfacial \ce{TaO_x} layer itself exhibiting a large dielectric constant ($\sim$7)\cite{lan_2023} while also acting as an effective platform facilitating the subsequent high-k \ce{HfO_2} growth. It has been shown, for example, that a \ce{TaO_x}/\ce{HfO_2} stack leads to significantly improved electrical characteristics compared to \ce{AlO_x}/\ce{HfO_2}, exhibiting very high on currents (up to 760 \textmu A/\textmu m) and a minimal subthreshold swing (70 mV/dec). \cite{lan_2023}

Even as the effect of the seeding layer on device performance is broadly accepted, the exact physical mechanisms linking this layer to \ce{MoS_2} transistor characteristics remain unclear.  In particular, the impact of seed layers on the effectiveness of charge transfer doping is still an open question.  To investigate, it is necessary to assess how variations in seed layers influence: (1) the TMD, (2) its doping \via charge transfer and (3) the quality of the resulting oxide stack (\ie seed and dielectric) so as to link the physical processes causing material changes to device performance. 

In response, we compared a series of Ta-seed layer depositions of varying thickness and purity during the fabrication of n-type monolayer (1L) \ce{MoS_2} FETs operating in enhancement mode. By comparing device characteristics in tandem with Raman, photoluminescence (PL) and X-ray photoelectron (XPS) spectroscopies, process induced variations in the TMD and dielectric could be directly linked to electrical behavior. Thinner Ta-seeds deposited under conditions minimizing adventitious oxygen led to higher on-currents and moblity along with lower threshold voltages.  XPS-measured shifts in the Hf 4f binding energy of the overlaid \hfo indicate that the seed layer strongly modifies the interfacial electrostatic environment governing charge transfer from the dielectric to the \ce{MoS_2}.  In tandem, strong correlations were found between the device characteristics and optical spectra indicating that the seed induces disorder within the TMD. In aggregate, the work not only highlights the central role of seed-layer deposition on TMD device performance but also emphasizes how multi-modal spectroscopy can serve as an ``in fab" tool capable of assessing process variability impacting TMD device performance.

\section*{Results and Discussion}
\begin{figure}[htbp]
\centering
\includegraphics[width=4 in]{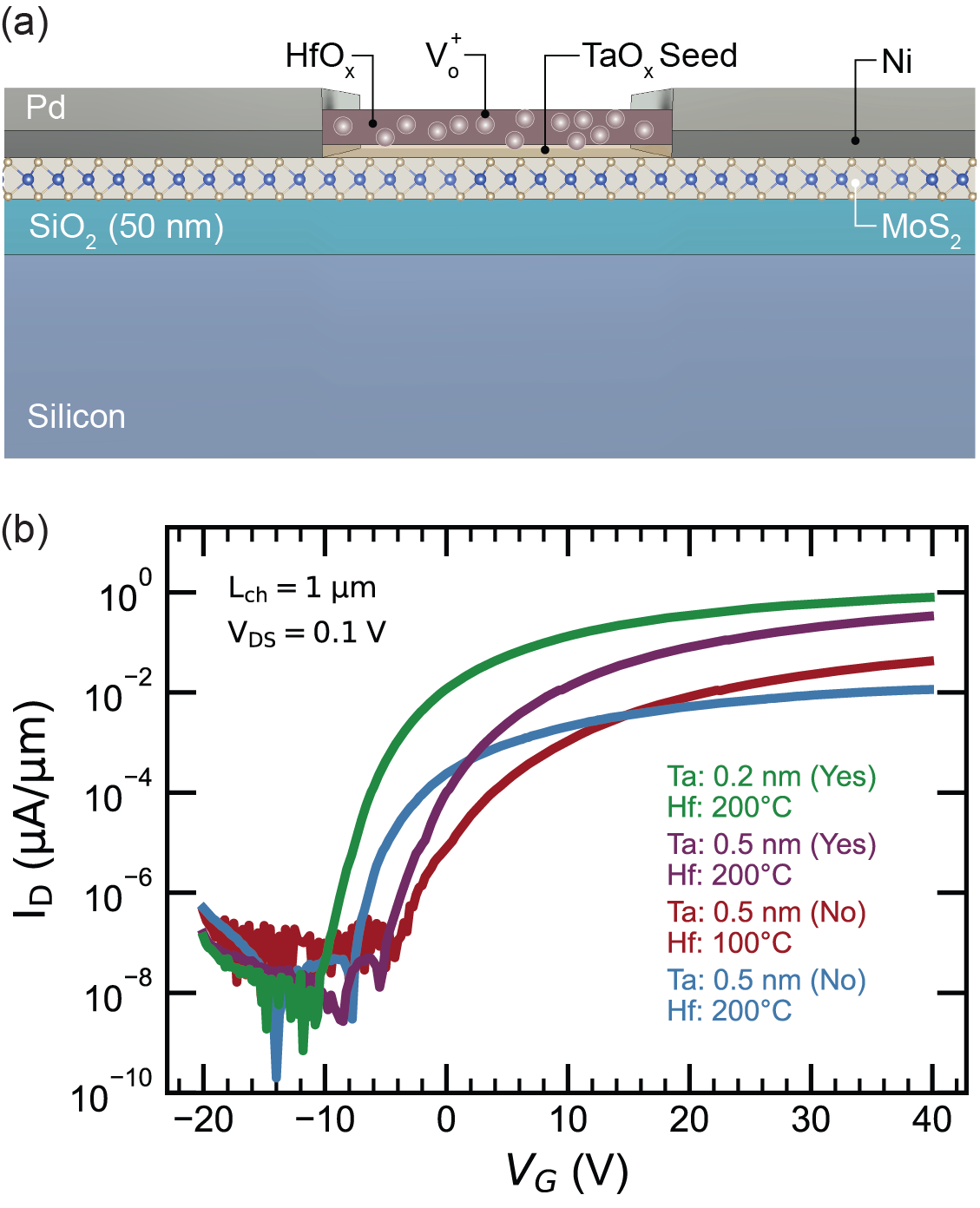}
\caption{(a) Schematic of the single-layer \ce{MoS_2} FETs employing a Ta-seed (\ie \tao) layer atop of which a 4.5-5 nm layer of \ce{HfO_x} was deposited for surface passivation. (b) Representative transfer curves for differing process splits in which varying the seed layer deposition significantly impacts device characteristics like \vth and \ion\!. These changes are attributed to both damage to the \ce{MoS_2} during seed-layer deposition and this layer's effect on the efficiency of charge transfer doping from oxygen vacancies in the \hfo and \tao dielectric stack.}
\label{Fig_1}
\end{figure}

As shown schematically in \fig{Fig_1}(a), monolayer n-type \ce{MoS_2} field-effect transistors were fabricated by transferring the TMD from a sapphire substrate to a Si/50 nm \ce{SiO_2} target wafer using a poly(methyl methacrylate)/thermal release tape (PMMA/TRT) based manual transfer method.\cite{schoenaers_2020}  Source/drain contacts were then patterned by a single electron beam (e-beam) lithography step, followed by deposition of Ni/Pd contacts and lift-off. Gating is applied through a body contact to the low-resistivity p-type silicon substrate as mediated by 50 nm of \ce{SiO_2}. Mechanical channel isolation is performed to minimize the process impact of chemical etching on \ce{MoS_2}. Devices had channel lengths spanning 0.5 to 20\mum\!. 

The top \ce{MoS_2} surface was passivated by deposition of a Ta-seed followed by hafnia.  The Ta-seed was deposited using e-beam evaporation at a rate of 0.1 or 0.2 \angstrom/s and then left to oxidize in ambient air for $>$ 24 hours.  Thicknesses for the Ta-seed reported hereafter correspond to the metal thickness prior to oxidation. Actual thicknesses of the \ce{TaO_x} layer were larger owing to the incorporation of oxygen. Subsequently, the terms Ta-seed and \tao are used synonymously to refer to the layer directly atop the \ce{MoS_2} within the device. After oxidation, a 4.5-5 nm layer of \ce{HfO_2} was deposited \via atomic layer deposition (ALD) at either 100°C or 200°C using \ce{Hf(N(CH_3)_2)_4} as the hafnium precursor. While the stoichiometry of this layer was not quantified explicitly, it is hereafter denoted as \ce{HfO_x} to highlight the slight oxygen deficiency typical of thin hafnia films.\cite{price_2019,beechem_2024a,vega_2025}

Four separate dielectric stacks were investigated that differed in: (1) Ta-seed thickness and deposition rate (0.2 nm $\&$ 0.1 \angstrom/s or 0.5 nm $\&$ 0.2 \angstrom/s) (2) pre-conditioning of the Ta-target before deposition (yes or no) and (3) temperature of the ALD process (100°C or 200°C).  Preconditioning refers to purposely evaporating from the Ta-target onto a ``blank" before initiating deposition on the device stack.   This step removes any oxide that may have formed on the Ta-target prior to its evaporation.\cite{schelfhout_2015}  Avoiding this step, therefore, is believed to increase the amount of oxygen present during Ta-seeding.  Note that while the \ce{TaO_x}/\ce{HfO_x} stack is envisaged as a high-K dielectric stack for gating, it was used here only as a passivation layer. This was to enable direct assessment of how deposition of this top-dielectric stack affected the characteristics of the \ce{MoS_2} channel divorced from any variable electrostatic capability in gating through these layers. 

Electrical measurements were performed using a FormFactor PAV 200 automated probe station under high vacuum conditions ($\sim 1\E{-4}$ Torr) and in the absence of light. A Keysight B1500A semiconductor parameter analyzer was used for data acquisition. Device parameters including threshold voltage (\vth\!), on-current (\ion\!), and field-effect mobility ($\mu$) were extracted from the transfer characteristics using a maximum transconductance ($g_m$) method. This method performs a linear fit to the transfer curve in the region of maximum transconductance and takes \vth to be the x-intercept of the fitted curve.\cite{cheng_2022c} The on-current was extracted at a fixed overdrive voltage of 10 V ($V_{G} - V_{TH} \, = 10 \mathrm{V}$). The two-terminal field-effect mobility was deduced through $g_m$ \via $\mu = (g_m \, L_{ch} ) / (C_{ox} \, V_{DS})$, where $L_{ch}$ is the channel length and $C_{ox}=690\;\mathrm{\frac{\mu F}{m^2}}$ the capacitance from the underlying \ce{SiO_2}. All reported measurements were acquired at a fixed drain bias of $V_{DS} = 0.1$ V.  SS was quantified by taking the maximum log-current slope within the drain current window of $1\E{-8}-1\E{-4}\;\mathrm{\frac{\mu A}{\mu m}}$. Hysteresis was quantified using a log-normalized current method described within the Methods Section. 

Devices were measured both before and after deposition of the overlaying Ta-seed and \ce{HfO_x} layers.  Electrical characteristics were collected for all 15 channel lengths (0.5 to 20 \mum\!) leading to the characterization of $\sim$20-80 devices for each process split. All traces are provided in Fig. S1 and S2 of the Supporting Information.  Subsequent box and whisker plots comparing the electrical characteristics include only devices with $L_{ch}\leq 10 \; \mu m$, however, as for some process splits yield of the longest-channel devices was very low. 
\begin{figure}[htbp]
\centering
\includegraphics[width=\textwidth]{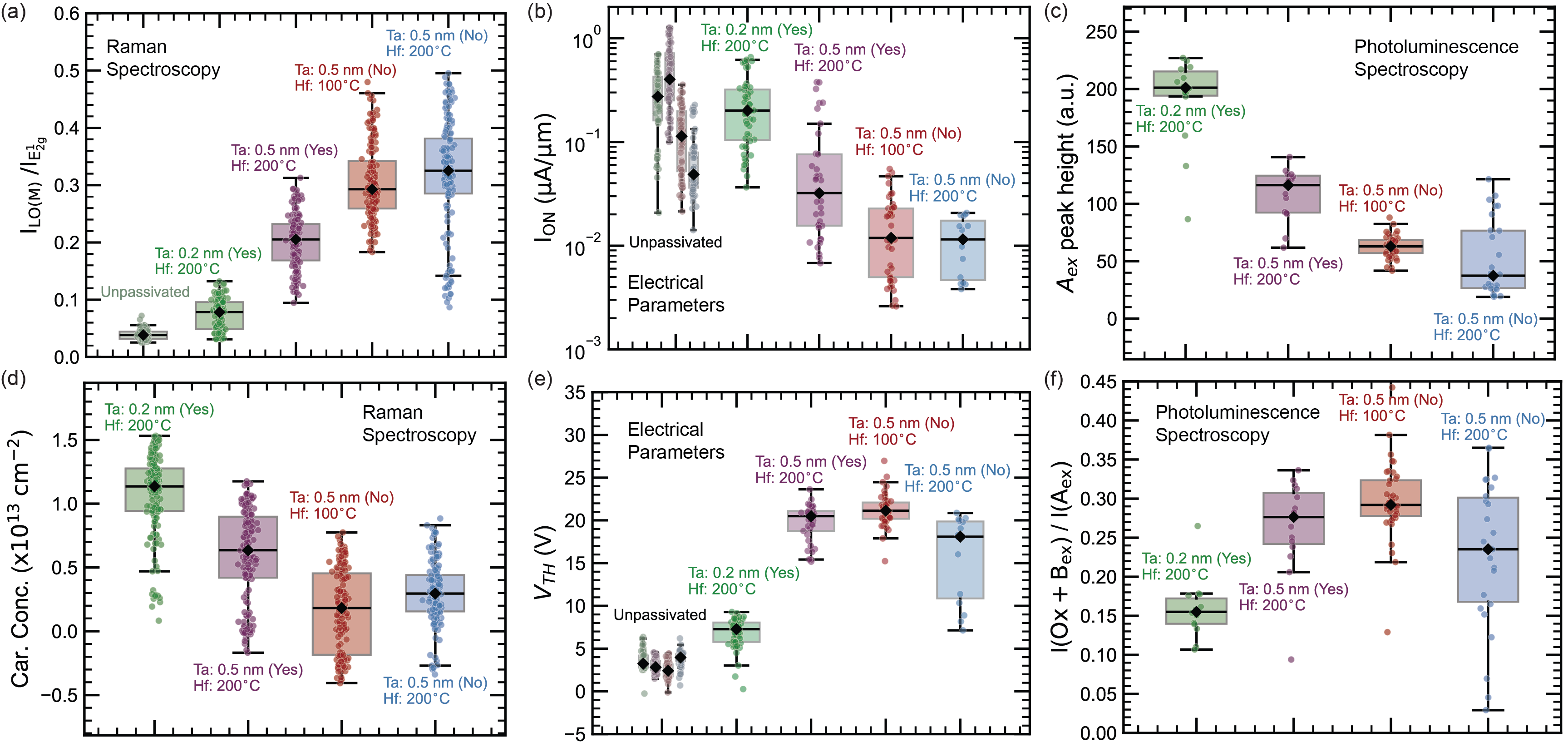 }
\caption{Correlation between fitted spectral and transistor characteristics. The (a) $I_{LO(M)}/I_{E^1_{2g}}$ Raman-ratio shows negative correlation with the (b) on-current (\ion\!) while the (c) PL-derived A-exciton peak height exhibits a similar trend relative to the process split. Similar negative correlation is observed when comparing Raman-derived (d) carrier concentration and (e) \vth while the (f) ratio of the two PL-features trends in a similar manner as the electrical characteristic. Together, the similarities between the electrical and spectral characteristics with process splits suggest differences in (a-c) TMD-disorder induced by the passivation and (d-f) the degree of charge transfer doping.}
\label{Fig_Box}
\end{figure}

Transistor performance changed significantly based upon how the Ta-seed and \ce{HfO_x} were deposited.  Threshold voltage (\vth\!) and on-current (\ion\!), in particular, were each heavily process dependent as can be seen qualitatively upon examination of the transfer curves in \fig{Fig_1}(b) for representative devices having a 1\mum channel length. Quantitatively, \fig{Fig_Box}(b,e) displays box and whisker plots of these parameters over all channel lengths before and after passivation.  Variation in the threshold voltage ($V_{TH}$) and on-current ($I_{on}$) was significantly smaller before passivation compared to that observed after deposition of the \tao/\hfo stack. Threshold voltages differed by over 10 V even as their pre-deposition values were within 0.5 V. On-current, meanwhile, reduced at greatly varying levels depending on the process even as they were comparable before passivation.   Taken together, it is readily apparent that the Ta-seed/\ce{HfO_x} depositions have affected the characteristics and quality of the devices. The following examines the physical mechanisms behind this variation by probing: (1) damage to the 2D-TMD material itself, (2) differences in the oxides deposited atop it and (3) the impact that these alterations have on charge transfer doping of the channel. 

To assess the physical changes in the materials inducing these variations, Raman and PL spectra of all 15 channel lengths were collected for each set of devices using a Witec alpha300R system with a 532 nm laser focused with a 50x/0.55 NA objective to a diffraction limited spot size of $\sim$590 nm nominally midway between the source and drain. Probing midway along the channel minimizes any effect due to charge-transfer or strain that could be induced by the contacts. Spectral measurements were acquired while all terminals of the device were allowed to float. Laser power and integration times for a given technique were kept consistent to allow nominal comparisons of signal strength between samples. Full details on the collection of Raman and PL spectra are provided in the Methods section. 

\begin{figure}[htbp]
\centering
\includegraphics[width=\textwidth]{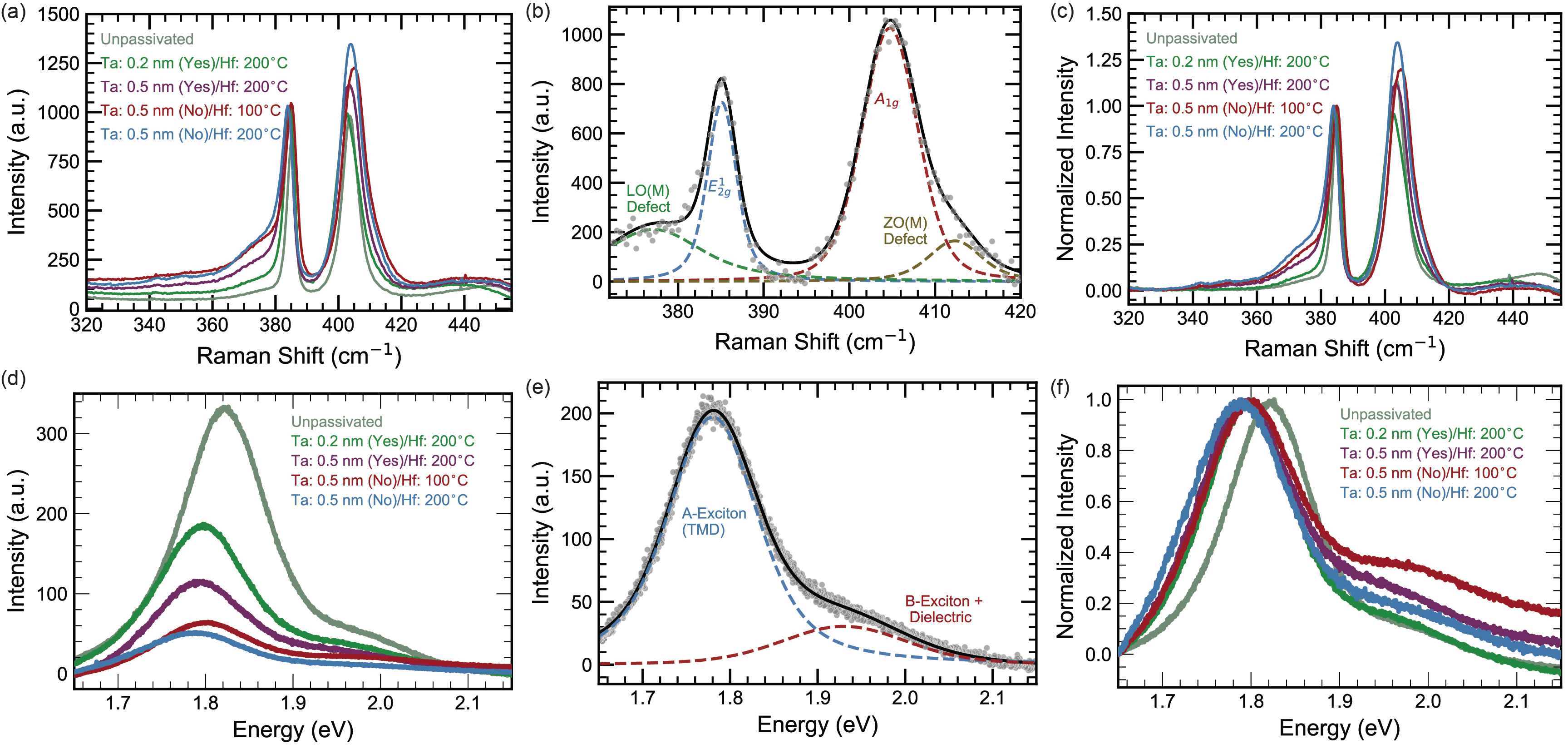}
\caption{ Average (a) Raman and (d) photoluminescence spectra for each of the differing Ta-seed/\ce{HfO_x} depositions along with the same spectra normalized to (c) the \ce{E^1_{2g}} Raman and (f) A-exciton PL features.  Qualitative differences emerge depending on deposition conditions for both methods.  Representative fitting for a (b) Raman and (e) PL-spectrum.  Characteristics of the fits are correlated with \ion and \vth\!.}
\label{Fig_Spectra}
\end{figure}

Average spectra acquired using both Raman and PL across process conditions are provided in \fig{Fig_Spectra} both in their ``as collected" [\fig{Fig_Spectra}(a,d)] and normalized [\fig{Fig_Spectra}(c,f)] forms. Differences in the spectral response of the samples are clear with the unpassivated spectra exhibiting the narrowest full-width at half-maximum (FWHM) of the Raman spectra and the largest overall PL-intensity compared to that of the fully fabricated devices. With passivation, a broadening of the Raman spectrum and a reduction in the PL-intensity occurs that varies depending on the process conditions. For example, the preconditioned, thin, Ta-seed layer (shown in green) has a Raman spectrum that is nominally similar to its unpassivated state.  The non-preconditioned, thicker, seeding layers (shown in red and blue) in contrast have a notable ``shoulder" in their Raman response between 360-380 \cm.   In addition, even as the total intensity of the PL-signal decreases in these thicker, non-preconditioned samples,  the relative strength of the PL-signal at higher energies ($>$ 1.9 eV) increases [\fig{Fig_Spectra}(f)]. Altogether, these spectral variations suggest that the gross differences in device performance arise because of differences in the TMD and/or Ta-seed/\ce{HfO_x} dielectric stack rather than the contacts since these spectral tools do not probe beneath the metal layers.  

To quantify these differences while also monitoring process induced variations in strain and doping, the Raman spectrum from each device of every process split was fit using a sum of four Voigt distributions.  In choosing the fitting procedure, note that while the characteristic \ce{MoS_2} \ce{E^1_{2g}} and \ce{A_{1g}} peaks at $\sim$382 \ce{cm^{-1}} and $\sim$407 \ce{cm^{-1}} are clearly visible, broadening is apparent to differing degrees near the lower and higher energy sides of the \ce{E^1_{2g}} and \ce{A_{1g}} modes, respectively. The broadening is associated with defect induced signals, with the \ce{E^1_{2g}} ``shoulder" resulting from the longitudinal optical (LO) mode near the M-point of the Brillouin zone [LO(M)], and the secondary peak near \ce{A_{1g}} originating from the corresponding out-of-plane optical mode [ZO(M)].\cite{ko_2016, mignuzzi_2015} Consequently, fitting was accomplished through the sum of four separate Voigt distributions corresponding to the first order \ce{E^1_{2g}} and \ce{A_{1g}} modes of \ce{MoS_2} and defect-induced signals arising from the LO(M) and ZO(M)-modes. A representative Raman spectrum\textemdash along with its corresponding fit\textemdash is shown in \fig{Fig_Spectra}(b).

Differences in disorder within the \ce{MoS_2} correlated with electrical performance as is apparent upon examination of \fig{Fig_Box}(a,b). Since the LO(M) mode becomes more clearly visible with increased disorder in the system \cite{nattoo_2025}, its intensity ratio to the characteristic \ce{E^1_{2g}} peak of the \ce{MoS_2} (\ie $\tfrac{I_{LO(M)}}{I_{E^1_{2g}}}$) is indicative of overall quality within the TMD itself. This ratio is plotted in \fig{Fig_Box}(a) for each process split where the disorder characteristic is negatively correlated with \ion [\fig{Fig_Box}(b)] and mobility (see Supporting Information Fig. S3).   Predictably, the best functioning devices (\ie largest \ion\!) exhibit the lowest defect signal in the Raman-intensity.  

Compared to disorder, strain had little, if any, impact on the electrical performance. Little correlation was found between \ion\!, for example, and the measured strain (see Supporting Information Fig. S4). Moreover, strain remained largely constant throughout the fabrication process.  Therefore, we conclude that observed differences in the device performance are not primarily driven by mechanical effects. 

Fitting of the photoluminescence spectra further supports the conclusion that variation in \ion is due primarily to disorder induced by the passivation.  Practically, PL-spectra were fit to the sum of two Voigt distributions nominally representing the A-exciton ($\sim$ 1.8 eV) along with a higher energy mode near 2 eV that we speculate could be influenced by not only the B-exciton energy of \ce{MoS_2} but also defects in the \tao and \hfo having PL-signatures in this energy range as well.\cite{nuytten_2025,ni_2008f,beechem_2024,zhu_2006a}  A representative fit is provided \fig{Fig_Spectra}(e). From these fits, a similar trend is found between the A-exciton intensity near 1.8 eV with \ion [compare \fig{Fig_Box}(b) and (c)]. The similarity is attributed to differing degrees of disorder within the TMD caused by the deposition of the \tao / \hfo. 

While PL-intensity could be reduced by either increasing carrier concentration within the \ce{MoS_2} or changing optical properties of the dielectric stack,\cite{berg_2019} trends in the Raman intensity and threshold voltage preclude such a deduction.  Similarities in the Raman intensity between samples [\fig{Fig_Spectra}(a)] indicate that optical properties cannot be the leading cause for the reduction in PL-intensity. Moreover, while the A-exciton intensity and \ion monotonically decreases in \fig{Fig_Box}(b,c), \vth ``rolls over" in moving from the non-preconditioned low-temperature (red) to high-temperature (blue) \hfo deposition. Since \vth is related to the amount of charge present in the TMD and the Raman derived carrier concentration\cite{iqbal_2020} also shows a similar ``roll over" [see \fig{Fig_Box}(d,e)],  the trion induced reduction in PL-intensity \cite{mak_2013} that comes with increasing charge concentration cannot, therefore, be the cause of the observed reduction. Instead, we conclude that increased non-radiative recombination associated with disorder is the primary cause of the reduction in PL-intensity.\cite{mccreary_2018} 

Disorder cannot be the cause for all variation in device behavior.  The ``roll over" seen in \vth and carrier concentration\textemdash as well as the ratio of PL-intensities [see \fig{Fig_Box}(f)]\textemdash suggest that something is affecting the amount of charge in the TMD beyond just structural defects and disorder. The defect signatures, in contrast, show a ``monotonic" response with process condition [compare \fig{Fig_Box}(a,b) with (d,e)]. Additional factors must be at play.  The following shows that the seed-layer impacts not only disorder in the TMD but also the effectiveness of charge transfer doping. 

\begin{figure}[htbp]
\centering
\includegraphics[width=\textwidth]{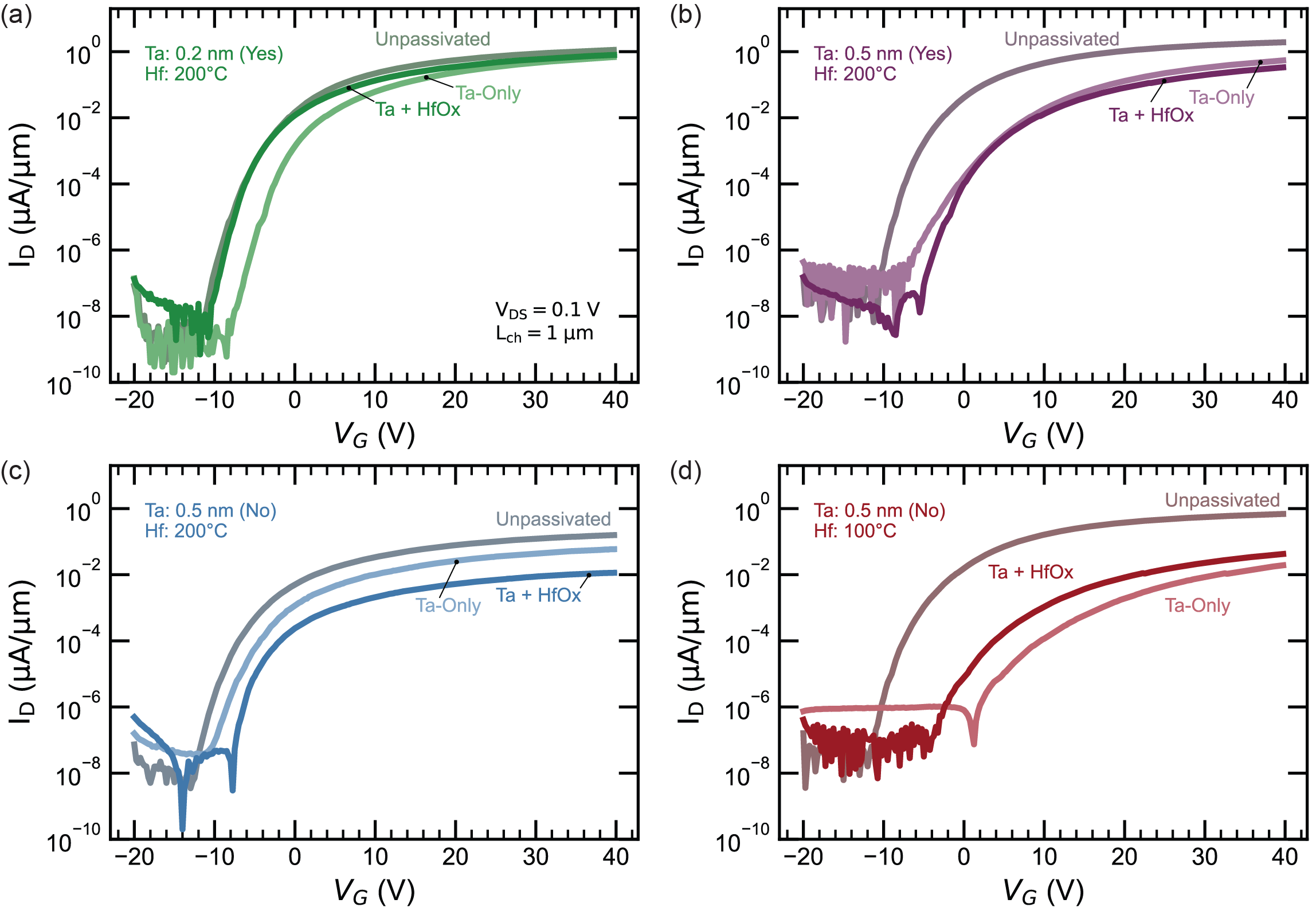}
\caption{Representative transfer curves for each process split comparing devices at each stage during fabrication of the \tao / \hfo stack, namely: unpassivated, covered with only the oxidized Ta-seed and then fully passivated with \tao + \hfo\!. Note the improvement in \ion that occurs with deposition of the \hfo for a subset of the process splits. All traces were for devices having a channel length of 1 \mum\!.}
\label{Fig_4}
\end{figure}

Transfer curves acquired throughout the fabrication process support this conclusion.  Upon deposition of the Ta-seed, \ion decreases and \vth increases irrespective of the process (see \fig{Fig_4}). This behavior is consistent with disordering, and possibly partial oxidation, of the \ce{MoS_2} as the Ta-seed is deposited and itself subsequently oxidizes. Somewhat counterintuitively, however, \hfo deposition partially restores performance for a subset of the process splits inducing an increase in \ion\!. As atomic force microscopy indicates nearly continuous \tao layers after even after the thinnest seed-layer deposition (see Supporting Information Fig. S5), it is unlikely that the \hfo ALD process is physically modifying the \ce{MoS_2} in a substantial way.  It is also noteworthy that hysteresis decreases relative to its unpassivated state in those devices exhibiting improvements in \ion with the \hfo deposition (see Supporting Information Fig. S6). Subthreshold swing, in contrast, was largely unaffected by the passivation (See Supporting Information Fig. S3).

The observed increase in \ion with \hfo deposition accompanied by the reduction in hysteresis\textemdash along with the associated variation in \vth and carrier concentration\textemdash together suggest that the \hfo is somehow acting to modify the charge environment  of the \ce{MoS2} even as it is not ``touching" the TMD.  The changes in device behavior caused by the \hfo deposition are occurring in a physically indirect manner.  Thus, these changes must be mediated by the Ta-seed. Charge transfer doping is inherently indirect and can occur as \ce{MoS_2} comes into contact with both \tao and \hfo\!.\cite{valsaraj_2015,lan_2023} Mechanistically, this occurs \via some combination of direct charge transfer from defects in the dielectric layers to the TMD or through virtual gating of the channel by the charged dipoles within the \tao/\hfo stack.  Thus, we conclude that the variation in \vth is a consequence of the process dependent degree to which charge transfer doping takes place.  

\begin{figure}[htbp]
\centering
\includegraphics[width=\textwidth]{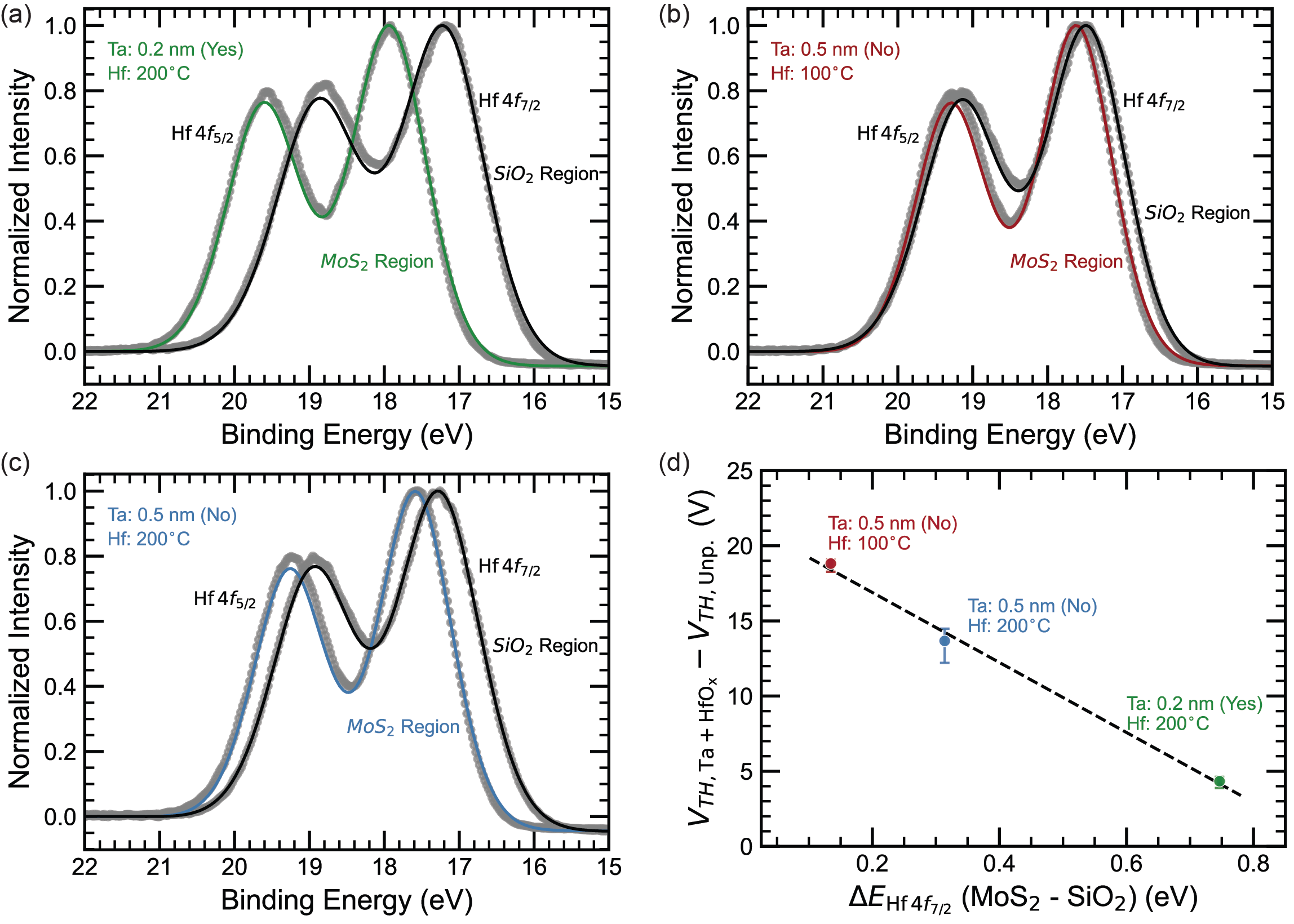}
\caption{XPS-spectra of \ce{HfO_x} obtained via collection of the \ce{Hf_{4f}} lines for samples having Ta-seed thicknesses of (a) 0.2 nm (preconditioned target, 200$^{\circ}$C \hfo\!) (b) 0.5 nm (no preconditioning, 100$^{\circ}$C \hfo\!) and (c) 0.5 nm (no preconditioning, 200$^{\circ}$C \hfo\!).  A spectrum was acquired where the \ce{HfO_x} overlapped the \ce{MoS_2} (\ce{MoS_2} region) and where the TMD was absent (\ce{SiO_2} region). (d) Correlation between shift in binding energy of $\mathrm{Hf}\,4f_{7/2}$ and \vth\!.}
\label{Fig_5}
\end{figure}

To support this assertion, X-ray photoelectron spectroscopy (XPS) examined the $\mathrm{Hf}\,4f_{5/2}$ and $\mathrm{Hf}\,4f_{7/2}$ features for devices exhibiting differing degrees of \hfo\!-induced recovery in \ion.  For each sample, an XPS-spectrum was collected from regions with, and without, \ce{MoS_2}. The resulting spectra are provided in \fig{Fig_5}, where a definitive shift in the XPS-spectra is apparent between the two locations irrespective of process. These shifts in the XPS spectra between regions with (\ce{MoS_2}-Region), and without (\ce{SiO_2}-Region) are highly-correlated with \vth\!, as shown by \fig{Fig_5}(d).  For example, the largest shift is observed for the preconditioned 0.2 nm Ta-seed [green, \fig{Fig_5}(a)] having the most negative \vth\!.  

A rigid-shift of the XPS spectrum is indicative of a change in the local charge environment.  The observed shifts are, therefore, signs of how much the charge environment differed in the \ce{HfO_x} for regions ``on" and ``off" the \ce{MoS_2}. Practically, we interpret the shift of the XPS spectrum to be caused by the transfer of charge from the \ce{HfO_x} and into the \ce{MoS_2} as would occur with charge transfer doping facilitated by oxygen vacancy defects.\cite{valsaraj_2015} As such, larger shifts correspond to greater charge transfer doping. Charge transfer doping, in turn, provides a knob to control threshold voltage. 

While the process dependent changes in the traces shown in \fig{Fig_4} highlight that the charge transfer doping primarily occurs with \hfo deposition, the Ta-seed plays a critical role in this process.  This is deduced empirically by observing that devices having nominally equivalent \hfo layers exhibited large variation in \vth (see \fig{Fig_Box}). By extension, the Ta-seed affects charge transfer doping. The resulting \tao itself must, therefore, be different depending on the process.  

XPS shows that the \tao does indeed vary depending on the deposition process.  Shown in \fig{Fig_6}, the oxidized Ta-seed exhibits clear spectral differences with a change in process conditions.  Changing the Ta-seed thickness (\ie 0.2 to 0.5 nm), target preconditioning, and deposition rate (0.1 to 0.2 \angstrom/s) broadens the $\mathrm{Ta}\,4f_{5/2}$ and $\mathrm{Ta}\,4f_{7/2}$ envelope with the emergence of a higher binding-energy doublet near 27 and 29 eV that is consistent with \ce{Ta_2O_5}.\cite{perez_2019, chen_1997b} Even so, a lower-energy doublet is observed in both samples, roughly 1 eV below the \ce{Ta_2O_5} response. This lower binding-energy feature indicates an additional Ta chemical environment that is distinct from stoichiometric \ce{Ta_2O_5}. Such a feature could arise from interfacial bonding of the Ta with the underlying \ce{MoS_2} (\eg in a mixed \ce{Ta_{2-x}Mo_xO_5} environment) and/or the presence of a sub-stoichiometric \tao region.  Regardless of the exact assignment, these XPS results indicate that the small alterations in Ta-seed deposition conditions appreciably alter the distribution of Ta chemical states and that these changes impact device characteristics.

\begin{figure}[htbp]
\centering
\includegraphics[width=4 in]{Figures/Fig_XPS_Ta_V3.png}
\caption{XPS-spectra of the $\mathrm{Ta}\,4f_{5/2}$ and $\mathrm{Ta}\,4f_{7/2}$ for devices undergoing separate Ta-seed deposition processes: (a) 0.2 nm, preconditioning, 0.1 \angstrom/s and (b) 0.5 nm no preconditioning, 0.2 \angstrom/s. The spectra exhibit two primary Ta chemical environments: a higher-binding-energy doublet consistent with \ce{Ta_2O_5}, and a lower-binding-energy doublet attributed to an interfacial and/or sub-stoichiometric \tao component.}
\label{Fig_6}
\end{figure}

It is clear that the seed-layer affects device performance because of its impact on charge transfer doping and must, therefore, be engineered in view of this process.  Practically, the seed layer can impact charge transfer doping through two primary mechanisms. First, it must facilitate charge transport from either itself or the \ce{HfO_x} layer. Beyond facilitating transport, however, the seed can also promote the presence of oxygen vacancies within the \ce{HfO_x} that are the primary source of charge that ultimately dopes the TMD-channel.  As sub-stoichiometric \ce{TaO_x} is known to take oxygen from \ce{HfO_x},\cite{jiang_2016,bednarz_2026} a more oxygen lean Ta-seed will likely result in more oxygen vacancies in the \ce{HfO_x} and thus greater charge transfer doping. Alternatively, higher oxygen content in the seed could mitigate this charge transfer. 

The link between seed layer differences, oxygen vacancies, and charge transfer doping motivates our speculation that the PL-signal at energies beyond 1.9 eV is not solely due to the B-exciton of \ce{MoS_2} but also defects in the overlaying \ce{TaO_x} and \ce{HfO_x}. Oxygen vacancies in \ce{HfO_x} produce a photoluminescence signal near the $\sim$2 eV energy of the B-exciton of \ce{MoS_2}.\cite{ni_2008f,beechem_2024a, vega_2025} Defects in tantalum oxide also exhibit PL in this energy range.\cite{zhu_2006a} Recognizing this\textemdash along with the observation that the ratio of the high to low-energy PL-modes correlate more with the threshold voltage (charge) rather than \ion (defects) [see \fig{Fig_Box}(d,e)]\textemdash leads us to conclude that the signal is influenced by the defects causing charge transfer doping as well as the B-exciton. Therefore, this PL-ratio may serve as a means of tracking charge transfer doping throughout the fabrication process. 

The interpretation presented here is consistent with the characteristics of devices made using Ta-seeds previously reported by a portion of our group.\cite{lan_2023}  Like that seen here, use of a Ta-seed led to an increase in on-current and shift to smaller threshold voltages when \hfo was deposited atop the seed. However, these same changes in electrical characteristics were also observed by Lan \etal \cite{lan_2023}\textemdash albeit to a lesser degree\textemdash after just the initial Ta-seed deposition. The seeming discrepancy occurs due to differences in architecture between the devices studied here and those of Lan \etal\! \cite{lan_2023} Previously, Ta-seeds were studied in a dual-gated arrangement in which the \ce{MoS_2} was not only overlaid by \tao / \hfo stack but also rested directly atop \hfo as well. In this arrangement, positively charged impurities within the \tao can electrostatically induce charge transfer doping from the bottom \hfo\!.  In the present study, \ce{SiO_2} sits beneath the \ce{MoS_2} instead. \ce{SiO_2} has a much lower defect concentration compared to \hfo and produces little, if any, charge transfer doping.  Thus, after Ta-seed deposition, TMD degradation wins out as there is no ``boost" stemming from charge transfer doping from the bottom dielectric.

\section*{Conclusion}
Widespread adoption of two-dimensional \ce{MoS_2} as an n-type transistor channel for scaled logic requires effective integration with the surrounding dielectric materials. The absence of out-of-plane bonding inherent to these van Der Waals solids complicates this integration and often necessitates the use of seed layers within the dielectric stack. Here, we show that such seed layers both induce disorder in the active channel and influence charge-transfer doping from the dielectric into the \ce{MoS_2}. Improved electrical performance is directly linked to thinner Ta seed layers deposited under oxygen-poor conditions, which limit deposition-induced damage while facilitating charge transfer. Finally, the strong correlation between spectral signatures and device characteristics highlights the practical utility of multimodal spectroscopy as a ``during-fab'' indicator of eventual device performance, enabling both process development and process monitoring.

\section*{Methods}
\textbf{Hysteresis Quantification}
Hysteresis was quantified by a log-normalized current criterion rather than a fixed value owing to the large variation in \ion observed among the process splits. For each forward and reverse sweep of the gate voltage, off ($I_{OFF}$) and on (\ion\!) currents were tabulated as the geometric mean between the two sweep directions. $I_{OFF}$ was taken to be the minimum measured current of the trace with a floor of $1\E{-9}\; \mathrm{\frac{\mu A}{\mu m}}$ applied before the logarithmic interpolation. The target current was defined as, $I_{\eta} = I_{\mathrm{OFF}}^{1-\eta}I_{\mathrm{ON}}^{\eta}$ where $\eta$ is a normalized log-current coordinate defined as \eqn{\eta =\frac{\log_{10}I_D-\log_{10}I_{\mathrm{OFF}}}{\log_{10}I_{\mathrm{ON}}-\log_{10}I_{\mathrm{OFF}}}}. Note that the off-current corresponds $\eta=0$ while $\eta=1$ corresponds to \ion\!. The signed hysteresis was calculated as $\Delta V_H(\eta)=V_{G,\mathrm{rev}}(I_{\eta})- V_{G,\mathrm{for}}(I_{\eta})$ at $\eta=0.5$, corresponding to the geometric midpoint of the device-specific current window. Similar trends in hysteresis were found when using $\eta=0.4-0.7$.  Finally, while the described method allows for quantitative comparisons of hysteresis, we acknowledge that for measurements where the subthreshold slope for the forward and backwards sweep are substantially different caution should be applied in utilization of the exact magnitudes. 

\textbf{Photoluminescence (PL) Measurements}
Photoluminescence spectra were acquired on a WITec Alpha 300R spectral imaging system with a 488 nm (2.54 eV) exciting laser irradiating the device at 0.42 mW across a diffraction limited area realized using a 50X/0.55 NA objective. The spectra were gathered using a UHTS 600 VIS spectrometer with a spectral resolution of less than 0.03 eV. For unpassivated films, all spectra were collected in a nitrogen environment using a THMS600 Linkam stage with a minimum 2-hour purge time. After passivation, measurements were acquired in ambient air. Powers were selected to minimize any changes in the materials induced by the laser-based characterization and were below levels known to induce significant thermal or photochemical alterations (see Supporting Information Fig. S7 and S8).  

\textbf{Raman Measurements}
Raman spectra were collected on a WITec Alpha 300R spectral imaging system with a 532 nm exciting laser irradiating the device at 0.42 mW across a diffraction limited area realized using a 50X/0.55 NA objective. The spectra were gathered using a UHTS 600 VIS spectrometer with a spectral precision of $\sim$0.016 \ce{cm^{-1}}.  Like in PL, unpassivated films were measured under a nitrogen purge and in ambient once covered in the \tao/\hfo stack. 

\textbf{X-ray Photoelectron Spectroscopy (XPS) Measurements}
XPS data were obtained using a Kratos Axis Ultra DLD spectrometer with monochromic Al K$\alpha$ radiation (1486.6 eV) at pass energy (PE) of 20 and 160 eV for high-resolution and survey spectra, respectively. A commercial Kratos charge neutralizer was used to avoid non-homogeneous electric charge of samples and to achieve better resolution. Typical instrument resolution for PE of 20 eV is $\sim$0.35 eV. Binding energy (BE) values refer to the Fermi edge and the energy scale was calibrated using $\mathrm{Au}\,4f_{7/2}$ at 84.0 eV and $\mathrm{Cu}\, 2p_{3/2}$ at 932.67 eV. Samples were placed on a stainless steel sample holder bar using a double-sided Cu tape. XPS data were analyzed with CasaXPS software. Prior to data analysis, the \ce{Si \, 2p} peak was set to a binding energy of 103.5 eV (corresponding to \ce{SiO_2} to correct for charge on each sample. Curve-fitting was performed following a Shirley background subtraction using Gaussian/Lorentzian peak shapes. The atomic concentrations of the elements in the near-surface region were estimated after a Shirley background subtraction taking into account the corresponding Scofield atomic sensitivity factors and inelastic mean free path (IMFP) of photoelectrons using standard procedures in the CasaXPS software, which assumes a homogeneous mixture of the elements within the collection depth of $\sim$10 nm.

\section{Supporting Information}
Supporting information providing: (1) transistor transfer curves for all examined devices in both logarithmic and linear scales, (2) quantification of mobilities and subthreshold swing (SS), (3) comparisons of strain throughout fabrication and between processes (4) atomic force microscopy images of the films after Ta-seed deposition and subsequent oxidation, (5) an examination of hysteresis, and (6) an analysis on the heating and effects of laser irradiation during optical spectroscopy.

\section*{Acknowledgements}
We acknowledge the joint MOU between the Indiana Economic Development Corporation, Purdue University, and imec, with in-kind support from the Applied Research Institute. D.L. acknowledges imec’s Industrial Affiliation program for exploratory logic. During preparation of this manuscript, the authors used ChatGPT for editorial assistance, including improving clarity and readability of selected text, and for assistance drafting and troubleshooting Python code used in data analysis. The tool was also used as a discussion aid when evaluating possible interpretations of the data. All final analyses, interpretations, code, and manuscript text were reviewed, edited, and approved by the authors, who take full responsibility for the content of the work.

\section*{Data Availability}
The data that support the findings of this study are available from the corresponding author upon reasonable request.


\providecommand{\latin}[1]{#1}
\makeatletter
\providecommand{\doi}
  {\begingroup\let\do\@makeother\dospecials
  \catcode`\{=1 \catcode`\}=2 \doi@aux}
\providecommand{\doi@aux}[1]{\endgroup\texttt{#1}}
\makeatother
\providecommand*\mcitethebibliography{\thebibliography}
\csname @ifundefined\endcsname{endmcitethebibliography}  {\let\endmcitethebibliography\endthebibliography}{}


\end{document}